\documentclass[pra,english,preprintnumbers,amsmath,amssymb,nofootinbib,longbibliography,twocolumn,superscriptaddress]{revtex4-1}
\usepackage[utf8]{inputenc}
\usepackage[T1]{fontenc}
\usepackage{amsmath}
\usepackage{amssymb}
\usepackage{graphicx}
\usepackage{tabularx}
\usepackage{bm}
\usepackage{pifont}
\usepackage{amsmath}

\DeclareMathOperator{\csch}{csch}

\usepackage{color}
\definecolor{orange}{rgb}{1,0.5,0}

\definecolor{bblue}{rgb}{0.2,0.8,1.0}

\newcommand{\beq}{\begin{equation}}
\newcommand{\eeq}{\end{equation}}
\newcommand{\bea}{\begin{eqnarray}}
\newcommand{\eea}{\end{eqnarray}}
\newcommand{\tr}{\mathrm{Tr}}

\begin{document}

\title{Virial coefficients of trapped and un-trapped three-component fermions \\ with three-body forces in arbitrary spatial dimensions}

\author{A. J. Czejdo}
\affiliation{Department of Physics and Astronomy, University of North Carolina, Chapel Hill, North Carolina 27599, USA}

\author{J. E. Drut}
\affiliation{Department of Physics and Astronomy, University of North Carolina, Chapel Hill, North Carolina 27599, USA}

\author{Y. Hou}
\affiliation{Department of Physics and Astronomy, University of North Carolina, Chapel Hill, North Carolina 27599, USA}

\author{J. R. McKenney}
\affiliation{Department of Physics and Astronomy, University of North Carolina, Chapel Hill, North Carolina 27599, USA}

\author{K. J. Morrell}
\affiliation{Department of Physics and Astronomy, University of North Carolina, Chapel Hill, North Carolina 27599, USA}

\date{\today}

\begin{abstract}
Using a coarse temporal lattice approximation, we calculate the first few terms of the virial expansion of a three-species 
fermion system with a three-body contact interaction in $d$ spatial dimensions,
both in homogeneous space as well as in a harmonic trapping potential of frequency $\omega$. Using the three-body problem to renormalize, 
we report analytic results for the change in the fourth- and fifth-order virial coefficients $\Delta b_4$ and $\Delta b_5$ as functions of
 $\Delta b_3$. Additionally, we argue that in the $\omega \to 0$ limit the relationship $b_n^\text{T} = n^{-d/2} b_n$ holds between
the trapped (T) and homogeneous coefficients for {\it arbitrary} temperature and coupling strength (not merely in scale-invariant regimes). 
Finally, we point out an exact, universal (coupling- and frequency-independent) relationship 
between $\Delta b_3^\text{T}$ in 1D with three-body forces and $\Delta b_2^\text{T}$ in 2D with two-body forces.
\end{abstract}

\maketitle 

\section{Introduction}

Motivated by the recent interest in one-dimensional (1D) Fermi and Bose gases in the fine-tuned situation where only three-body 
interactions are present~\cite{PhysRevA.97.061603, PhysRevA.97.061604, PhysRevA.97.061605, PhysRevA.97.011602, PhysRevA.99.053607, 
PhysRevA.99.013615, PhysRevA.100.013614, PhysRevA.100.063604,Daza:2018nvg}, we explore here the thermodynamics of fermions 
with a contact three-body interaction in the region of low fugacity (which corresponds to a dilute regime and therefore high temperatures in units of the energy scale set by the density). We focus on the fermionic case but explore the problem in arbitrary dimension $d$. To that end, we implement a semiclassical lattice approximation 
(SCLA) to calculate the virial coefficients $b_n$, and carry out their evaluation up to $n = 5$ at leading order (LO) in that approximation.

The LO-SCLA was introduced in Ref.~\cite{PhysRevA.98.053615} as a way to estimate virial coefficients in two-component Fermi gases. The approximation seems crude in its definition but performs surprisingly well when
the lowest non-trivial order in the virial expansion is used as a renormalized coupling constant ($b_2$ for two-body forces,
for example, and $b_3$ in this work). Not surprisingly, the approximation was seen to work better at weak coupling, which
makes sense as the radius of convergence of the virial expansion was found to be quickly reduced as a result of the interaction.
In Ref.~\cite{PhysRevA.100.063627}, the NLO-SCLA was explored up to $b_7$, displaying the convergence properties up to the unitary 
point (in 3D) and in Ref.~\cite{PhysRevA.100.063626} the LO-SCLA was used for systems in a harmonic trap, showing that the approximation 
can capture the dependence on the trap frequency $\omega$. In both cases, the analytic dependence of virial coefficients on the dimension was obtained, as will be the case here. This is to be contrasted with conventional methods to calculate virial coefficients, which can be very precise but are limited to specific situations (coupling strength, dimension, etc.) and are typically unable to provide analytic insight as they are entirely numerical.

Our analytic formulas for the virial coefficients, although approximate, support and shed light on the relationship $b_n^\text{T} \to n^{-d/2} b_n$ in the $\omega \to 0$ limit, where the superindex T indicates the harmonically trapped situation.
This connection is well-known to be valid in the noninteracting limit and in the so-called unitary limit of spin-$1/2$ fermions in 3D, both 
of which
feature temperature-independent coefficients $b_n$. As we will argue, that relationship is actually valid for all temperatures and 
coupling constants, and holds for three-body interactions just as well as for two-body interactions.
Finally, we point out an exact, coupling- and frequency-independent relationship between the $\Delta b_3^\text{T}$ in 1D with three-body forces and $\Delta b_2^\text{T}$ in 2D with two-body forces.

\section{Hamiltonian and virial expansion} 

We focus on a non-relativistic Fermi system with a three-body contact interaction, such that the 
Hamiltonian for three flavors $1,2,3$ is $\hat H = \hat T + \hat V$, where
\bea
\label{Eq:T}
\hat T \!=\! {\int{d^d x\,\hat{\psi}^{\dagger}_{s}({\bf x})\left(-\frac{\hbar^2\nabla^2}{2m}\right)\hat{\psi}_{s}({\bf x})}}
\eea
and
\bea
\label{Eq:V}
\hat V \!=\! - g_{d}\! \int{d^d x} \,\hat{n}_{1}({\bf x})\hat{n}_{2}({\bf x})\hat{n}_{3}({\bf x}),
\eea
where the field operators $\hat{\psi}_{s}, \hat{\psi}^{\dagger}_{s}$ are fermionic fields for particles of type $1,2,3$ (summed over $s$ above), and $\hat{n}_{s}({\bf x})$ are the coordinate-space densities. In the remainder of this work, we will take $\hbar = k_\text{B} = m = 1$. Besides the above, we will also consider the case in which an external trapping potential term is added to 
the Hamiltonian, of the form
\beq
{\hat V}_\text{ext} \!=\! \frac{1}{2}m \omega^2 \! \int{d^d x}\; {\bf x}^2 \left[ \hat n_1({\bf x}) + \hat n_2({\bf x})+\hat n_3({\bf x}) \right].
\eeq

One way to characterize the thermodynamics is through the virial expansion~\cite{LIU201337}, which is an expansion around the dilute limit
$z\to 0$, where $z=e^{\beta \mu}$ is the fugacity, i.e. it is a low-fugacity expansion. The corresponding coefficients
accompanying the powers of $z$ in the expansion of the grand-canonical potential $\Omega$ are the virial coeffiecients;
specifically,
\beq
-\beta \Omega = \ln {\mathcal Z} = Q_1 \sum_{n=1}^{\infty} b_n z^n
\eeq
where 
\beq
\mathcal Z = \tr \left[e^{-\beta (\hat H - \mu \hat N)}\right] = \sum_{N=0}^{\infty} z^N Q_N
\eeq 
is the grand-canonical partition function, $Q_1$ is the one-body partition function, $b_1 = 1$, and the higher-order 
coefficients require solving the corresponding few-body problems:
\bea
Q_1 b_2 &=& Q_2 - \frac{Q_1^2}{2!},\\
Q_1 b_3 &=& Q_3 - b_2 Q_1^2  - \frac{Q_1^3}{3!},\\
Q_1 b_4 &=& Q_4 -  \left(b_3 + \frac{b_2^2}{2}\right) Q_1^2 -b_2\frac{Q_1^3}{2!}  - \frac{Q_1^4}{4!},\\
Q_1 b_5 &=& Q_5 - (b_4  + b_2 b_3 ) Q_1^2 - \left (b_2^2  + b_3 \right )\frac{Q_1^3}{2} \nonumber \\
&&- b_2 \frac{Q_1^4}{3!}  - \frac{Q_1^5}{5!}, 
\eea
and so forth.

Since $Q_1 \propto V$, the above expressions display precisely how the volume dependence cancels out in each $b_n$.
In particular, the highest power of $Q_1$ will always involve single-particle (i.e. noninteracting) physics and will therefore cancel 
in the change due to interactions $\Delta b_n$, such that
\bea
Q_1 \Delta b_2 &=& \Delta Q_2 \\
Q_1 \Delta b_3 &=& \Delta Q_3 - \Delta b_2 Q_1^2 ,\\
Q_1 \Delta b_4 &=& \Delta Q_4 -  \Delta\left(b_3 + \frac{b_2^2}{2}\right) Q_1^2 -\frac{\Delta b_2}{2} Q_1^3, \\
Q_1 \Delta b_5 &=&\Delta  Q_5 - \Delta (b_4  + b_2 b_3 ) Q_1^2 \nonumber \\
&&- \frac{1}{2}\Delta \left (b_2^2  + b_3 \right )Q_1^3 - \frac{\Delta b_2}{3!} Q_1^4,
\eea
and so on. Note that, when only three-body interactions are present, as is the case we consider here, there is no change in the two-body
spectrum, i.e. $\Delta b_2 = 0$. Therefore, the above expressions simplify to
\bea
Q_1 \Delta b_3 &=& \Delta Q_3,\\
Q_1 \Delta b_4 &=& \Delta Q_4 -  \Delta b_3 Q_1^2, \\
Q_1 \Delta b_5 &=&\Delta  Q_5 - \left(\Delta b_4  + b_2 \Delta b_3\right) Q_1^2 - \frac{\Delta b_3}{2} Q_1^3.
\eea

In terms of the partition functions $Q_{MNL}$ of $M$ particles of type 1, $N$ of type 2, and $L$ of type $3$, we have
\bea
\Delta Q_3 &=& \Delta Q_{111}, \\
\Delta Q_4 &=& 3\Delta Q_{211}, \\
\Delta Q_5 &=& 3\Delta Q_{311} + 3\Delta Q_{221}.
\eea
From the above equations we see that there is only a small number of non-trivial contributions to each virial coefficient. The main task is calculating each of these terms and for that purpose we use a coarse lattice (or semiclassical) approximation, as explained next.

\section{The semiclassical approximation \\ at leading order}

To carry out our calculations of virial coefficients we introduce a Trotter-Suzuki (TS) 
factorization of the Boltzmann weight. In the lowest possible order, the TS factorization
amounts to keeping only the leading term in the following formula:
\beq
e^{-\beta (\hat T + \hat V)} = e^{-\beta \hat T}e^{-\beta \hat V}\times e^{-\frac{\beta^2}{2} [\hat T, \hat V]}\times \dots,
\eeq
where higher orders involve exponentials of nested commutators of $\hat T$ with $\hat V$. Taking the leading order in this
expansion is equivalent to setting $[\hat T , \hat V] = 0$, which is why we refer to it as a semiclassical approximation.
As Refs.~\cite{PhysRevA.98.053615, PhysRevA.100.063626, PhysRevA.100.063627} have shown, this seemingly crude 
approximation provides surprisingly good answers, especially at weak coupling, and is therefore useful toward examining
the virial expansion in an analytic fashion. Below, we give two explicit examples of the application of our approximation
to the calculation of virial coefficients.

\subsection{A simple example: $\Delta b_3$}

As the simplest example, we consider $Q_{111}$:
\bea
Q_{111} &=& \sum_{{\bf p}_j} \langle {\bf P} | e^{-\beta \hat T}e^{-\beta \hat V} | {\bf P} \rangle \\
&&\!\!\!\!\!\!\!\!\!\!\!\! = \sum_{{\bf p}_j} e^{-\beta (p_1^2 + p_2^2 + p_3^2)/2m} \langle {\bf P}  |e^{-\beta \hat V} | {\bf P} \rangle,
\eea
where we have used a collective momentum index ${\bf P} = ({\bf p}_1,{\bf p}_2,{\bf p}_3)$.
Inserting a coordinate-space completeness relation 
to evaluate the potential energy factor, we obtain
\bea
e^{-\beta \hat V} | {\bf X} \rangle &=& \prod_{\bf z} (1 + C\hat{n}_{1}({\bf z})\hat{n}_{2}({\bf z}) \hat{n}_{3}({\bf z}))| {\bf X} \rangle  \\
&&\!\!\!\!\!\!\!\!\!\!\!\! = | {\bf X} \rangle + C \sum_{\bf z}\delta({\bf x}_1 - {\bf z}) \delta({\bf x}_2 - {\bf z}) \delta({\bf x}_3 - {\bf z})| {\bf X} \rangle \nonumber \\
&&\!\!\!\!\!\!\!\!\!\!\!\! = \left[1 + C \delta({\bf x}_1 - {\bf x}_3) \delta({\bf x}_2 - {\bf x}_3) \right]| {\bf X} \rangle \nonumber,
\eea
where $C = (e^{\beta g_{d}} - 1)\ell^{2d}$, $\ell$ is an ultraviolet regulator in the form of a spatial lattice spacing, 
and we used the fermionic relation $\hat{n}^2_s = \hat{n}_s$. We also introduced a collective index ${\bf X} = ({\bf x}_1,{\bf x}_2,{\bf x}_3)$. The $C$-independent term yields the noninteracting 
result, such that we may write
\bea
\label{Eq:Q111}
\Delta Q_{111} &=& C \sum_{{\bf p}_j ,{\bf x}_k} \!\!\! e^{-\beta (p_1^2 + p_2^2 + p_3^2 )/2m} \nonumber \\
&&\times \delta({\bf x}_1 - {\bf x}_3)\delta({\bf x}_2 - {\bf x}_3) |\langle {\bf X} | {\bf P} \rangle |^2,
\eea
which simplifies substantially when using a plane wave basis since $|\langle {\bf X} | {\bf P} \rangle |^2 = 1/V^3$,
where $V$ is the $d$-dimensional volume of the system. We then find
\beq
\Delta Q_{111} = C \frac{Q_{100}^3}{V^2} 
\eeq
where
\beq
Q_{100} = \sum_{{\bf p}_1} e^{-\beta p_1^2/2m}.
\eeq
Thus, 
\beq
\label{Eq:Db3vsC}
\Delta b_3 = C \frac{{Q_{100}^3}}{V^2 Q_1} = C \frac{Q_{1}^2}{27 V^2} = \frac{C}{\lambda_T^{2d}} \frac{1}{3}, 
\eeq
where $Q_1 = 3 Q_{100} = 3V / \lambda_T^d$, $\lambda_T = \sqrt{2 \pi \beta}$ is the 
thermal wavelength, and $V$ is the system's spatial volume. This relationship between
the bare coupling constant $C$ and the physical quantity $\Delta b_3$ provides a way to
renormalize the problem.
In other words, $\Delta b_3$ will play the role of the renormalized dimensionless coupling constant.

The general form of the change $\Delta Q_{MNL}$ in the partition
function for $M$ type-1 particles, $N$ type-2 particles and $L$ type-3 particles, with a contact interaction, is given by
\beq
\Delta Q_{MNL} = \sum_{{\bar {\bf P}},{\bar {\bf X}} } e^{-\beta {\bar {\bf P}}^2/2m} |\langle {\bar {\bf X}} | {\bar {\bf P}}\rangle|^2
(C f_a(\bar {\bf X}) + C^2 f_b(\bar {\bf X}) + \dots),
\eeq
where ${\bar {\bf P}},{\bar {\bf X}}$ represent all momenta and positions of the $M+N+L$ particles, and the
functions $f_a$, $f_b$, $\dots$, which encode the matrix element of $e^{-\beta \hat V}$, depend on the specific 
$MNL$ case being considered. The wavefunction $\langle {\bar {\bf X}} | {\bar {\bf P}}\rangle$ is a product of three Slater 
determinants which, if using a plane-wave single-particle basis, leads to Gaussian integrals over the momenta $\bar {\bf P}$. 

\subsection{Another example: $\Delta b_4$ in a harmonic trap.}

In this section we consider the case in which the system is held in a harmonic trapping potential of frequency $\omega$.
As the expressions for the virial coefficients in terms of the canonical partition functions carry over to this case, we will
simply add the superindex `T' to denote quantities in the trapped system.
To calculate $\Delta b_4^\text{T}$ we need $\Delta b_3^\text{T}$ and $Q_1^\text{T}$. The latter is of course trivial as there is no interaction
in that case (see Ref.~\cite{PhysRevA.100.063626}):
\bea
Q_1^\text{T}&=&3\sum_{\bf n}e^{-\beta E_{\bf n}}=3e^{-\beta \omega d /2}\bigg(\frac{1}{1-e^{-\beta \omega}}\bigg)^d\\ 
&=& 3\left( \frac{1}{2 \sinh(\beta \omega /2)} \right)^{d},
\eea
where $E_{\bf n}$ is the single-particle energy level of the harmonic oscillator (separated in $d$-dimensional cartesian coordinates
such that $\bf n$ represents a $d$-dimensional vector of harmonic oscillator quantum numbers).

To obtain $\Delta b_3^\text{T}$, we proceed as in the previous example to obtain the analogue of Eq.~(\ref{Eq:Q111})
for the trapped case:
\bea
\Delta Q^\text{T}_{111} &=& C \sum_{{\bf n}_j ,{\bf x}_k} \!\!\! e^{-\beta (E_{{\bf n}_1} + E_{{\bf n}_2} + E_{{\bf n}_3})} \nonumber \\
&&\!\!\!\!\!\!\! \times \delta({\bf x}_1 - {\bf x}_3)\delta({\bf x}_2 - {\bf x}_3) |\langle {\bf x}_1  {\bf x}_2 {\bf x}_3 | {\bf n}_1 {\bf n}_2 {\bf n}_3  \rangle |^2.
\eea
The sums over ${\bf x}_3, {\bf x}_2$ can be carried out right away, and moreover
\beq
|\langle {\bf x}_1  {\bf x}_2 {\bf x}_3 | {\bf n}_1 {\bf n}_2 {\bf n}_3  \rangle |^2 = 
|\phi_{{\bf n}_1}({\bf x}_1)|^2 |\phi_{{\bf n}_2}({\bf x}_2)|^2 |\phi_{{\bf n}_3}({\bf x}_3)|^2,
\eeq
where $\phi_{{\bf n}}({\bf x})$ is the single-particle harmonic oscillator wavefunction in $d$-dimensional cartesian coordinates.
Using the above, we obtain
\bea
\Delta Q^\text{T}_{111} &=& C \sum_{{\bf x}} \rho^3({\bf x}; \beta\omega),
\eea
where
\beq
\rho({\bf x};\beta\omega) = \sum_{{\bf n}} e^{-\beta E_{{\bf n}}} |\phi_{{\bf n}}({\bf x})|^2.
\eeq
Note that $\sum_{\bf x} \rho({\bf x};\beta\omega) = Q_1^\text{T} / 3$.

Using the Mehler kernel (see Ref.~\cite{PhysRevA.100.063626}) evaluated at equal spatial arguments, we find that
\beq
\rho({\bf x};\beta\omega) = \omega^{\frac{d}{2}} \frac{e^{- \omega \tanh(\beta \omega / 2) {\bf x}^2}}{\left( 2 \pi \sinh(\beta \omega)\right)^{\frac{d}{2}}},
\eeq
where we note that $\tanh(\beta \omega / 2) > 0$ for all $\beta \omega > 0$. 
Carrying out the resulting Gaussian integrals and simplifying,
\beq
\Delta b_3^\text{T} = \frac{\Delta Q^{T}_{111}}{Q_1^\text{T}} = \frac{C}{\lambda_T^{2d}} \frac{1}{3^{\frac{d}{2} + 1} } 
\left(\frac{\beta \omega}{\sinh(\beta \omega)}\right)^{d},
\eeq
where $\lambda_T = \sqrt{2 \pi \beta}$.

Note that, as $\beta \omega \to 0$, we obtain
\beq
\Delta b_3^\text{T} = \frac{C}{\lambda_T^{2d}} \frac{1}{3^{\frac{d}{2} + 1} } = \frac{1}{3^\frac{d}{2}} \Delta b_3,
\eeq
where in the last equality we have used Eq.~(\ref{Eq:Db3vsC}).

For $\Delta b_4^\text{T}$, we need $\Delta Q^\text{T}_{211}$, which is easily seen to be given by
\bea
\Delta Q^\text{T}_{211} &=& C\sum_{{\bf x},{\bf x'}} \rho^2({\bf x};\beta\omega) \left[\rho({\bf x};\beta\omega) \rho({\bf x}';\beta\omega) - 
\rho^2({\bf x},{\bf x}';\beta\omega)\right], \nonumber \\
&=& \Delta Q^\text{T}_{111} Q_1^\text{T}/3 - C\sum_{{\bf x},{\bf x'}} \rho^2({\bf x};\beta\omega) \rho^2({\bf x},{\bf x}';\beta\omega),
\eea 
where
\beq
\rho({\bf x},{\bf x}';\beta\omega) = \sum_{{\bf n}} e^{-\beta E_{{\bf n}}} \phi_{{\bf n}}({\bf x}) \phi_{{\bf n}}({\bf x}'),
\eeq
which, using the Mehler kernel, becomes
\beq
\rho({\bf x},{\bf x}';\beta\omega) =  \frac{\omega^{\frac{d}{2}} e^{- \omega \coth (\beta \omega) ({\bf x}^2 + {\bf x}'^2)/2 + \omega  \csch(\beta \omega) {\bf x}\cdot {\bf x}' }}{\left( 2 \pi \sinh(\beta \omega)\right)^{\frac{d}{2}}}.
\eeq

Thus, in the continuum limit,
\bea
\Delta b_4^\text{T} &=& 3\frac{\Delta Q^{T}_{211}}{Q_1^\text{T}} -  \Delta b_3^\text{T} Q_1^\text{T}  \nonumber \\
&=& 
- \frac{3C}{Q_1^\text{T}}\sum_{{\bf x},{\bf x'}} \rho^2({\bf x};\beta\omega) \rho^2({\bf x},{\bf x}';\beta\omega) \nonumber \\
&=& -\frac{C}{\lambda_T^{2d}} \frac{1}{2^{\frac{d}{2}}}
\left[
\frac{\beta \omega }{\sinh(\beta \omega)} 
\frac{1}{(1 + 3 \cosh(\beta \omega))^{\frac{1}{2}}}
\right]^{d} \\
&=& -\frac{3^{\frac{d}{2}+1}}{2^{\frac{d}{2}}} \frac{1}{(1 + 3 \cosh(\beta \omega))^{\frac{d}{2}}} \Delta b_3^\text{T} .
\eea

Note that, in the $\beta \omega \to 0$ limit, our approximation yields
\bea
\label{Eq:Db4TbwZeroLimit}
\Delta b_4^\text{T} &=& 
- \frac{3^{\frac{d}{2}}}{2^{d}}  \frac{3}{2^{\frac{d}{2}}} \Delta b_3^\text{T} = 
- \frac{1}{2^{d}}  \frac{3}{2^{\frac{d}{2}}} \Delta b_3,
\eea
which we will use below.

\section{Results in homogeneous space}

\subsection{Virial coefficients}
Using the steps outlined above, we have calculated $\Delta b_4$ and $\Delta b_5$ and
obtained
\bea
\label{Eq:Db5Db5SCLA}
\Delta b_4 &=& -C \frac{Q_{1} Q_1(2\beta)}{9 V^2} = -3 \frac{Q_1(2\beta)}{Q_{1} } \Delta b_3 , \\
\Delta b_5 &=&  C \left ( \frac{\left(Q_1(2\beta)\right)^2}{9 V^2} +  \frac{Q_1 Q_1(3\beta)}{9 V^2}\right) \nonumber \\
&=&   \left ( \frac{3 \left(Q_1(2\beta)\right)^2}{Q_1^2} +  \frac{3 Q_1(3\beta)}{Q_1}\right) \Delta b_3,
\eea
for the fermionic three-species system with a three-body contact interaction in $d$ spatial dimensions. In the last equation,
the first term on the right-hand side represents the contribution of $Q_{221}$, and the second term that of $Q_{311}$.

In the continuum limit, it is easy to perform the resulting Gaussian integrals that determine
$Q_1$ and obtain
\bea
\label{Eq:Db4Db5SCLA}
\Delta b_4 &=& - \frac{3}{2^{\frac{d}{2}}} \Delta b_3 , \\
\Delta b_5 &=& 3\left ( \frac{1}{2^d} +  \frac{1}{3^{\frac{d}{2}}}\right) \Delta b_3.
\eea

Using these results, one may calculate the pressure, density, compressibility and even Tan's contact (with knowledge of $\Delta b_3$
as a function of the interaction strength, e.g. $\beta \epsilon_B$ in 1D or 2D, where $\epsilon_B$ is the trimer binding energy).
To provide a description of the thermodynamics that is as universal as possible across spatial dimensions, we will use 
$\Delta b_3$ as the measure of the interaction strength and display our results in terms of that parameter. Furthermore, one may
also define a (dimensionless) contact density as 
\beq
\label{Eq:ContactDb3}
\mathcal C = \frac{\lambda_T^{d}}{V}\frac{\mathcal \partial \ln \mathcal Z}{\partial \Delta b_3},
\eeq
which differs from the conventional definition by a chain-rule factor ${\partial \Delta b_3}/{\partial \lambda}$ (which in turn can be determined
by solving the three-body scattering problem), where $\lambda$ is the $d$-dimensional coupling constant. To make the
expression dimensionless, we have used the thermal wavelength $\lambda_T = \sqrt{2 \pi \beta}$.

\subsection{Thermodynamics and contact across dimensions}

The interaction change in the pressure $\Delta P$ can be written in dimensionless form 
in arbitrary dimension as
\beq
\beta V \Delta P = Q_1 \sum_{k=1}^{\infty} \Delta b_k \, z^k.
\eeq
Similarly, the interaction change in the density can be written as
\beq
\lambda_T^d \Delta n= 3 \sum_{k=1}^{\infty}  k \,\Delta b_k\,  z^k,
\eeq
and, using our definition of the contact in Eq.~(\ref{Eq:ContactDb3}),
\beq
\Delta \mathcal C = 3 \sum_{k=1}^{\infty} \frac{\partial \Delta b_k}{\partial \Delta b_3} \, z^k.
\eeq
\begin{figure}[b]
\includegraphics[width=1\columnwidth]{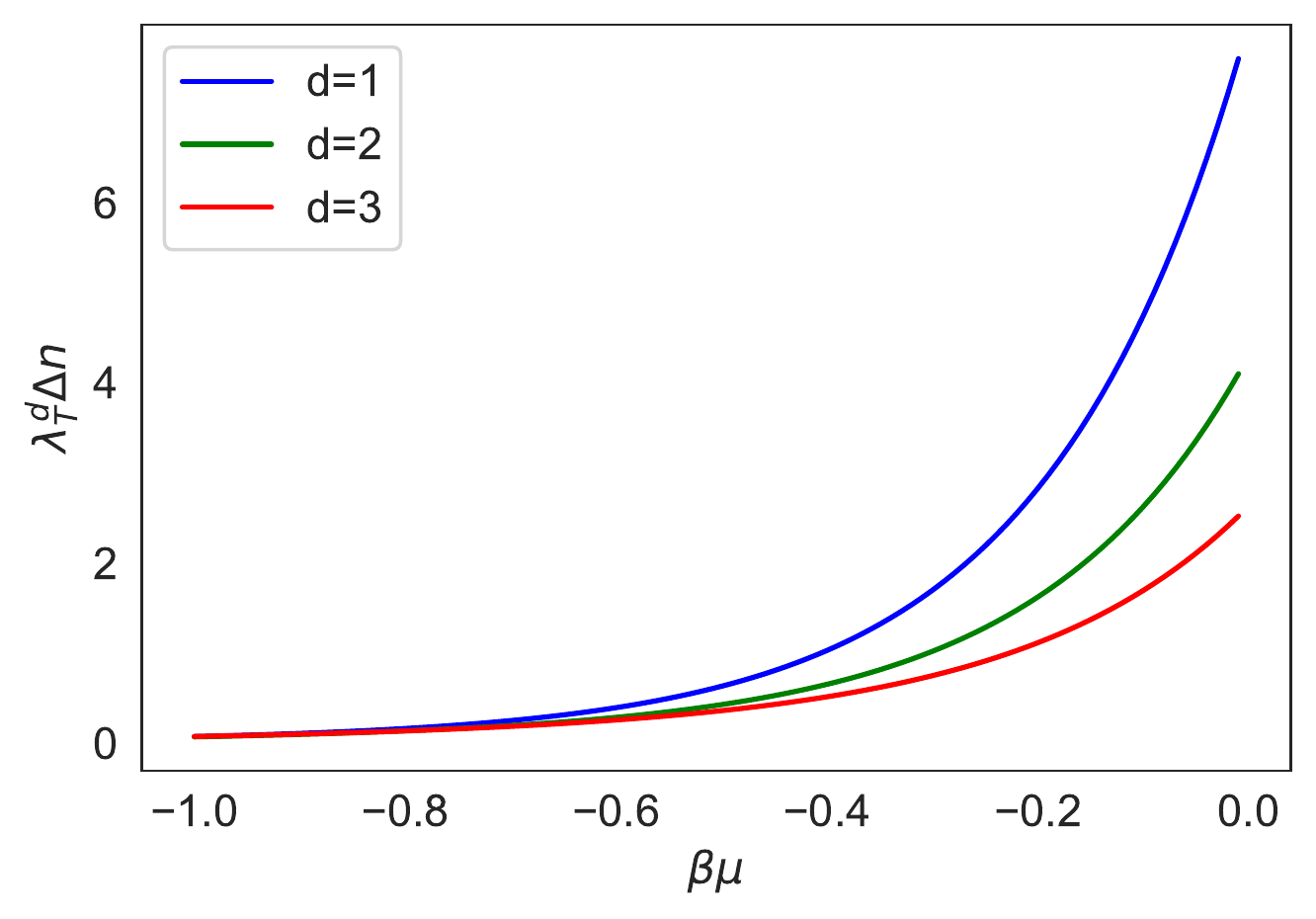}
\caption{\label{Fig:Density}
Density, in units of $\lambda^d_T = (2\pi \beta)^{d/2}$ as a function of $\ln z = \beta \mu$, at $\Delta b_3 = 0.25$.
}
\end{figure}
Implementing our LO-SCLA results, we obtain
\bea
\beta \lambda_T^d \Delta P &\simeq& 3 \Delta b_3 \, z^3\!\! \left[1  - \frac{3}{2^{\frac{d}{2}}} z + 3\left(\! \frac{1}{2^d} + \frac{1}{3^{\frac{d}{2}}}\!\right)z^2 \right] ,\\
\lambda_T^d \Delta n &\simeq& 9 \,\Delta b_3\, z^3\!\! \left[1 - \frac{4}{2^{\frac{d}{2}}}  z + 5 \left(\! \frac{1}{2^d} + \frac{1}{3^{\frac{d}{2}}}\!\right) z^2\right],\\
\Delta \mathcal C &\simeq&3z^3 \left[1 - \frac{3}{2^{\frac{d}{2}}} z + 3\left(\! \frac{1}{2^d} + \frac{1}{3^{\frac{d}{2}}}\!\right) z^2 \right].
\eea

As an example, in Fig.~\ref{Fig:Density} we display the density as a function of the logarithm of the fugacity $\ln z = \beta \mu$ for
$\Delta b_3 = 0.25$ and for $d=1,2,3$.

The behavior of $\Delta n$ as a function of $\beta \mu$ in Fig.~\ref{Fig:Density} is as expected for a system with attractive interactions,
namely the interaction-induced change in the density is positive and enhanced by increasing $\beta \mu$ (or, equivalently,
washed out at low densities, i.e. for large and negative $\beta \mu$). Also as expected (and as observed in Refs.~\cite{PhysRevA.98.053615}
and~\cite{PhysRevA.100.063627} for two-body interactions), interaction effects are more pronounced in lower dimensions at fixed $\Delta b_3$.

\section{Results in a harmonic trap}

\subsection{Fourth- and fifth-order virial coefficients}
\begin{figure}[b]
\includegraphics[width=1\columnwidth]{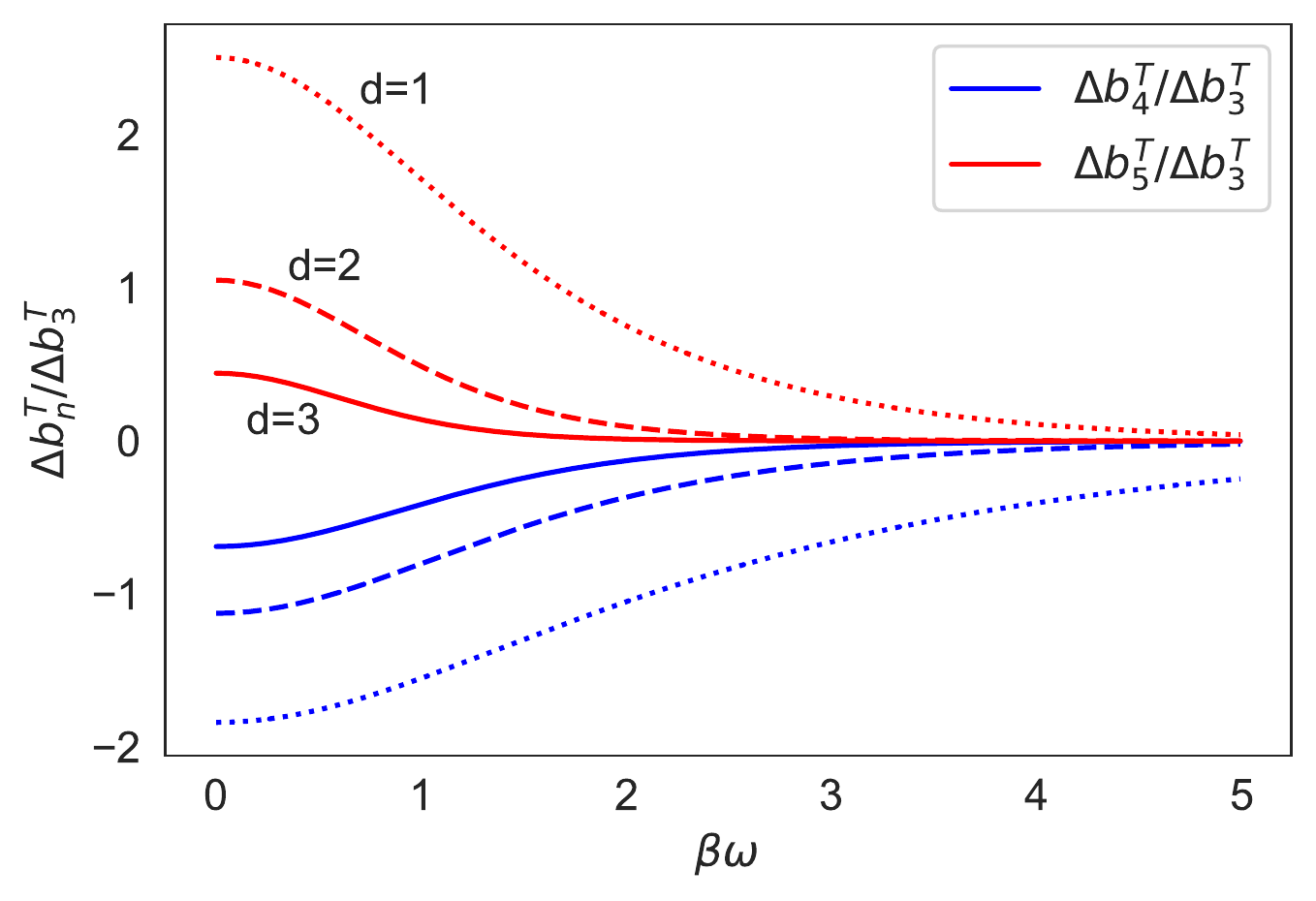}
\caption{\label{Fig:Db4TDb5ToverDb3T} $\Delta b_4^T$ (blue lines) and $\Delta b_5^T$ (red lines), in units of $\Delta b_3^T$, as a function of $\beta \omega$ in the LO-SCLA. Results are shown in $d=1$ (dotted), $d=2$ (dashed), and $d=3$ (solid).
}
\end{figure}
We have generalized our example of $\Delta b_4^\text{T}$, discussed in a previous section, to $\Delta b_5^\text{T}$.
For future reference, we show both results:
\bea
\label{Eq:Db4trap}
\Delta b_4^\text{T} &=& -\frac{3^{\frac{d}{2}+1}}{2^{\frac{d}{2}}} \frac{1}{(1 + 3 \cosh(\beta \omega))^{\frac{d}{2}}} \Delta b_3^\text{T},
\eea
\bea
\label{Eq:Db5trap}
\Delta b_5^\text{T} &=&  
3^{\frac{d}{2}+1}\left ( \left[ \frac{1}{12\cosh^2(\beta\omega)+4\cosh(\beta\omega)-1} \right]^{\frac{d}{2}} \right. \nonumber \\
&+& \left. \left[ \frac{1}{12\cosh^2(\beta\omega)+8\cosh(\beta\omega)} \right]^\frac{d}{2} \right) \Delta b_3^\text{T}.
\eea
In Fig.~\ref{Fig:Db4TDb5ToverDb3T} we show these results in $d=1,2,3$ as a function of $\beta \omega$.
In contrast to the behavior of $\Delta b_4^T$ for the case of two-body interactions, explored in Refs.~\cite{PhysRevLett.116.230401, PhysRevA.100.063626},
here both $\Delta b_4^\text{T}$ and $\Delta b_5^\text{T}$ display monotonic behavior.
Furthermore, at this order in the SCLA, both $\Delta b_4^\text{T}$ and $\Delta b_5^\text{T}$ are proportional to $\Delta b_3^\text{T}$, such that
the results of Fig.~\ref{Fig:Db4TDb5ToverDb3T} are universal predictions in the sense of being coupling-independent.

\subsection{A universal relation in the $\beta \omega \to 0$ limit}

Note that, in the $\beta \omega \to 0$ limit, where the homogeneous system is recovered,
\beq
\label{Eq:Db5TbwZeroLimit}
\Delta b_5^\text{T} \to
3^{\frac{d}{2}+1}
\frac{1}{5^{\frac{d}{2}}}  \left (\frac{1}{2^d} + \frac{1}{3^{\frac{d}{2}}} \right) \Delta b_3^\text{T} = 
\frac{3}{5^{\frac{d}{2}}}  \left (\frac{1}{2^d} + \frac{1}{3^{\frac{d}{2}}} \right) \Delta b_3
\eeq

Using Eqs.~(\ref{Eq:Db4TbwZeroLimit}), (\ref{Eq:Db4Db5SCLA}), and (\ref{Eq:Db5TbwZeroLimit}), we find that trapped and un-trapped virial coefficients are related, in the $\beta \omega \to 0$ limit, as follows:
\bea
\Delta b_3^\text{T} &=& 3^{-\frac{d}{2}} \Delta b_3, \\
\Delta b_4^\text{T} &=& 4^{-\frac{d}{2}} \Delta b_4, \\
\Delta b_5^\text{T} &=& 5^{-\frac{d}{2}} \Delta b_5.
\eea
Although we have only explored $\Delta b_n^\text{T}$ for $n=3,4,5$ here (the cases $n=1,2$ are trivially satisfied as well), 
the fact that the above relationship holds points us to conjecture that the relation
\beq
\left . b_n^\text{T} \right |_{\beta \omega \to 0} = n^{-\frac{d}{2}} b_n,
\eeq
is universally valid for all $n$, couplings, and temperatures (it is well known to be satisfied by 
noninteracting gases). Other authors, see 
e.g.~\cite{PhysRevLett.102.160401, PhysRevA.82.023619, LIU201337} have noted (and proven using the local density approximation) that 
this relationship is satisfied in the unitary limit (where the $b_n$ are temperature-independent), and
the same connection was found for $n=3,4$ in systems with two-body forces in Ref.~\cite{PhysRevA.100.063626}
for arbitrary couplings (within the LO-SCLA). In principle, there is no special reason why $b_n^\text{T}$
should not approach $b_n^\text{}$ when the trapping potential is removed. That there is a $d$- and $n$-dependent
factor connecting those two quantities in the noninteracting case is merely a geometrical artifact of the choice of basis in which the 
calculations are performed (namely the harmonic oscillator basis in the trapped case and plane waves in the homogeneous case),
which has no impact on physical quantities.
Based entirely on dimensional analysis, however, the natural guess is that $b_n^\text{T}$ may approach $b_n^\text{}$
times a dimensionless function of temperature and other dynamical scales. [That would actually change the partition function
in a non-trivial way, in particular concerning Tan's contact, but let us put that aside for the moment.]
Such a dimensionless function could only result from the interplay between
the trapping potential ${\hat V}_\text{ext}$ and the interaction $\hat V$, possibly leading to subtleties in the $\omega \to 0$ limit 
(similar to those arising from degenerate perturbation theory).
However, the fact that $[{\hat V}_\text{ext}, \hat V] = 0$ suggests that there should be no such subtlety and therefore no residual 
dependence on interaction-related scales in the relationship between $b_n^\text{T}$ and $b_n$ as $\beta \omega \to 0$. 
In that limit, the dimensionless quantities $b_n^\text{T}$ and 
$b_n$ should be related by a coupling- and temperature-independent function; their connection should be entirely geometrical and 
fully determined by the noninteracting case, for which $b_n^\text{T} = n^{-\frac{d}{2}} b_n$ when $\beta \omega \to 0$. 
We therefore conclude that the conjecture is true for all $n$, coupling strengths, and temperatures.

\subsection{An exact relation across systems and dimensions}

Finally, we point out a coupling-independent relationship between the 1D case with a three-body interaction 
(i.e. the 1D case of the system studied in this work) and the 2D case with only two-body interactions (denoted below by the superindex ``2b2D'').
As pointed out in Ref.~\cite{PhysRevLett.120.243002}, there exists an exact relationship between the three-body problem of the former 
situation and the two-body problem of the latter. That relationship yields a simple proportionality rule between the
corresponding virial coefficients, given by
\beq
\Delta b_3 = \frac{Q_{111}^\text{cm}}{Q_{11}^\text{cm, 2b2D}} \frac{Q_1^\text{2b2D}}{Q_1} \Delta b_2^\text{2b2D},
\eeq
where the superscript ``cm'' indicates the partition function associated with the center-of-mass motion, which is not affected by the interactions
and completely factorizes (both in the spatially homogeneous as well as in the harmonically trapped case). 
In the spatially homogeneous case, the proportionality factor between $\Delta b_3$ and $\Delta b_2^\text{2b2D}$ is 
$1/\sqrt{3}$, as shown in Ref.~\cite{PhysRevLett.120.243002}. On the other hand, in the harmonically trapped case, the relationship becomes 
\beq
\Delta b_3^\text{T} = \frac{2}{3} \Delta b_2^\text{T,2b2D}.
\eeq
We stress that while this relationship is restricted to the 1D $\Delta b_3^\text{T}$ and $\Delta b_2^\text{T,2b2D}$, it is valid for all couplings and all 
values of $\beta \omega$ and is in that sense universal.

For completeness and future reference, we provide here details on the origin of this correspondence for the trapped case.
The Schr\"odinger equation for this system takes the form
\beq
\left[-\frac{\nabla^2_{\bf r}}{2m} + 
\frac{g}{m} \delta(x-y)\delta(y-z) +
\frac{1}{2}m\omega^2 {\bf r}^2 \right]
\psi({\bf r})
=
E \psi({\bf r})
\eeq
where $x$, $y$, and $z$ again indicate the different-flavor particles, ${\bf r} = (x,y,z)$, and
\beq
\nabla^2_{\bf r} = \frac{\partial^2}{\partial x^2}\!+\! \frac{\partial^2}{\partial y^2} \!+\! \frac{\partial^2}{\partial z^2}.
\eeq
Factoring out the center-of-mass (c.m.) motion by defining $Q = \frac{1}{\sqrt{3}}(x+y+z)$, $q_1 = \frac{1}{\sqrt{2}}(y-x)$, $q_2 = \frac{1}{\sqrt{6}}(x + y - 2z)$, 
and $\psi(x,y,z) = \Phi(Q)\phi({\bf q})$, with ${\bf q} = (q_1,q_2)$, we obtain
\beq
\left[ -\frac{1}{2m} \frac{\partial^2}{\partial Q^2} +\frac{1}{2}m\omega^2 Q^2 \right]\Phi(Q) = E_{\mathrm{c.m.}}\Phi(Q),
\eeq
for the c.m. motion, and
\beq
\left[-\frac{\nabla^2_{\bf q}}{2m}  + \frac{\tilde{g}}{m} \delta({\bf q})+ \frac{1}{2}m\omega^2 {\bf q}^2\right]\phi({\bf q}) = E_r \phi({\bf q}),
\eeq
where $\displaystyle \tilde{g} = {g}/{\sqrt{3}}$ is the effective coupling and $E_r$ is the energy of relative motion,
which is identical to that of a single particle in a 2D harmonic oscillator potential with a $\delta$-potential at the origin. 
This establishes the exact relationship between our three-body 1D problem and its two-body counterpart in 2D with two-body 
interactions.


As in the spatially homogeneous case, the eigenvalues $\epsilon_{\omega} = E_r/\omega$ of the
harmonically trapped system are determined implicitly, in this case as solutions to
\beq
\label{Eq:renorm}
\frac{1}{\tilde{g}} = \frac{1}{\pi} \sum_{n=0}^{\Lambda_{\omega}} \frac{1}{\epsilon_{\omega} - (2n+1)} \rightarrow 
\frac{1}{2\pi}\left[\psi_0\left(\frac{1-\epsilon_{\omega}}{2}\right) - \ln{\Lambda_{\omega}}\right],
\eeq
where $\psi_0(z)$ is the digamma function, where $\Lambda_{\omega}$ is a UV cutoff. Unlike in the untrapped problem, with its unique bound state, the trapped problem admits 
an infinite set of discrete excited states (all with positive energy). The problem is renormalized by relating the bare coupling to the 
$\epsilon_{\omega}$ occurring in the lowest energy branch.


\section{Summary and Conclusions}

In this work we have calculated the high-temperature thermodynamics of three-flavored Fermi gases
with a contact three-body interaction in $d$ spatial dimensions, as determined by the virial expansion.
We carried out calculations in homogeneous space as well as in a harmonic trapping potential of frequency $\omega$.
To that end, we implemented a coarse temporal lattice approximation at leading order (the LO-SCLA) and calculated
the change in the virial coefficients $\Delta b_n$ due to interaction effects. In that context, we established a relation between 
the first two non-trivial virial coefficients, namely $\Delta b_4$ and $\Delta b_5$, as functions of $\Delta b_3$.
In addition, we argued that in the $\beta \omega \to 0$ limit, the relationship $\Delta b_n^\text{T} = n^{-d/2} \Delta b_n$ holds between
the trapped and homogeneous coefficients for arbitrary $n$, coupling strengths, and temperatures; furthermore, it is valid for
systems with two- and three-body interactions. We showed that our calculations
reproduce that relationship for $n=3,4,5$. Finally, we showed a relationship between the harmonically trapped case in 1D with three-body 
interactions and its analogue in 2D with two-body interactions, namely $\Delta b_3^\text{T} = \frac{2}{3} \Delta b_2^\text{T,2b2D}$.


\acknowledgments
This material is based upon work supported by the National Science Foundation under Grant No.
PHY{1452635} (Computational Physics Program).

\bibliography{ThreeBody}

\end{document}